\documentclass[authoryear,final,1p,times]{elsarticle}

\usepackage{amssymb}
\usepackage{amsmath}

\begin{document}

\begin{frontmatter}

\title{Improving prediction models by incorporating external data with weights based on similarity}

\author[imbi,fdm]{Max Behrens}
\author[imbi,fdm,genmed]{Maryam Farhadizadeh}
\author[math]{Angelika Rohde}
\author[onco,leip]{Alexander Rühle}
\author[leip]{Nils H. Nicolay}
\author[imbi,fdm]{Harald Binder}
\author[imbi,fdm]{Daniela Zöller}

\affiliation[imbi]{organization={Institute of Medical Biometry and Statistics, Faculty of Medicine and Medical Center, University of Freiburg},
            addressline={Stefan-Meier-Straße 26}, 
            city={Freiburg},
            postcode={79104}, 
            country={Germany}}

\affiliation[fdm]{organization={Freiburg Center for Data Analysis and Modeling, University of Freiburg},
            addressline={Ernst-Zermelo-Straße 1}, 
            city={Freiburg},
            postcode={79104}, 
            country={Germany}}

\affiliation[math]{organization={University of Freiburg, Department of Mathematical Stochastics},
            addressline={Ernst-Zermelo-Straße 1},
            city={Freiburg},
            postcode={79104}, 
            country={Germany}}

\affiliation[onco]{organization={Department of Radiation Oncology, University of Freiburg–Medical Center},
            addressline={Robert-Koch-Straße 3},
            city={Freiburg},
            postcode={79106}, 
            country={Germany}}

\affiliation[leip]{organization={Department of Radiation Oncology, University of Leipzig},
            addressline={Stephanstr. 9a},
            city={Leipzig},
            postcode={04103}, 
            country={Germany}}

\affiliation[genmed]{organization={Institute of General Practice / Family Medicine, Faculty of Medicine and Medical Center, University of Freiburg},
            addressline={Elsässerstraße 2n}, 
            city={Freiburg},
            postcode={79110}, 
            country={Germany}}

\begin{abstract}
In clinical settings, we often face the challenge of building prediction models based on small observational data sets. For example, such a data set might be from a medical center in a multi-center study. Differences between centers might be large, thus requiring specific models based on the data set from the target center. Still, we want to borrow information from the external centers, to deal with small sample sizes. There are approaches that either assign weights to each external data set or each external observation. To incorporate information on differences between data sets and observations, we propose an approach that combines both into weights that can be incorporated into a likelihood for fitting regression models. Specifically, we suggest weights at the data set level that incorporate information on how well the models that provide the observation weights distinguish between data sets. Technically, this takes the form of inverse probability weighting. We explore different scenarios where covariates and outcomes differ among data sets, informing our simulation design for method evaluation. The concept of effective sample size is used for understanding the effectiveness of our subgroup modeling approach. We demonstrate our approach through a clinical application, predicting applied radiotherapy doses for cancer patients. Generally, the proposed approach provides improved prediction performance when external data sets are similar. We thus provide a method for quantifying similarity of external data sets to the target data set and use this similarity to include external observations for improving performance in a target data set prediction modeling task with small data.
\end{abstract}


\end{frontmatter}


\section{Introduction}

Clinical prediction is increasingly reliant on models derived from observational data, which presents unique challenges. Unlike data from controlled experiments or clinical trials, there often is not just a single, clearly defined data set. The analysis data set is a potentially small subgroup of a larger data set, e.g., when using data from a specific medical center where the prediction model is sought in the context of a multi-center data set. Such inherent divisions within the data entail further difficulties, such as differences in patient demographics, treatment protocols, and other factors \citep{madiganEvaluatingImpactDatabase2013, glynnHeterogeneityIntroducedEHR2019}. 
To address these challenges, we propose a method for improving prediction performance when incorporating similar data from external subgroups, e.g., other medical centers, for building a model for the target subgroup of interest. Specifically, observations of external subgroups are assigned weights that include individual and subgroup-specific components. These components are based on a propensity score approach \citep{rosenbaumCentralRolePropensity1983,austinMovingBestPractice2015} that estimates the probability of originating from the target subgroup. 

The idea of either subgroup-level weights or individual observation weights in prediction models has been investigated in the literature in different contexts. For example, \cite{weyerWeightingApproachJudging2015} considered high-dimensional molecular data and introduced a weighted regression approach for deriving gene expression signatures for a specific subgroup of a data set. There, all observations in a subgroup receive the same weight, and these weights are systematically varied to visualize the effect on the resulting gene expression signatures. Building on this, \cite{richterModelbasedOptimizationSubgroup2019} propose to optimize the subgroup weights with respect to prediction performance. In contrast, \cite{madjarWeightedCoxRegression2021} propose to use an individual weight for each observation. Specifically, they developed a weighted Cox regression, using a similarity measure from \cite{bickelMultitaskLearningHIV2008} to create a sample that matches the target subgroup's distribution. Interestingly, a combination of both approaches, i.e., subgroup-specific weights that reflect information shared by all observations in the data set and observation-specific weights that recognize the variability within external data sets, has not been considered so far.  

As an alternative approach, a global model across all subgroups could be considered. For example, mixed effects models \citep{pinheiro2006mixed} could be used for modeling heterogeneity across data sets and observations. If differences can be explained by some continuous covariates, varying coefficient models could be used \citep{hastieVaryingCoefficientModels1993, fanStatisticalEstimationVarying1999}. However, such modeling comes with additional assumptions. For mixed-effects models, it is assumed that there exists a combination of fixed effects, which remain constant across data sets, and random effects, used to model heterogeneity and are typically considered to follow specific distributions. Varying coefficient models might impose even stronger assumptions, such as smooth variation of effects. 

We want to avoid such assumptions and instead focus on weighting approaches. Our focus is on investigating the potential benefit of combining subgroup-specific and observation-specific weights. The latter is obtained as the probability of membership of the target subgroup estimated by a regression model. The subgroup-specific weight is based on the prediction performance of this regression model, as strong prediction performance indicates that the target data set and the external data set can be distinguished well, i.e., are dissimilar. Specifically, the area under the curve (AUC), which can be interpreted as the probability of a random pair of observations from the two data sets, is combined with an inverse probability approach.  

\begin{figure}[tbh]
\begin{center}
\includegraphics[width=\textwidth]{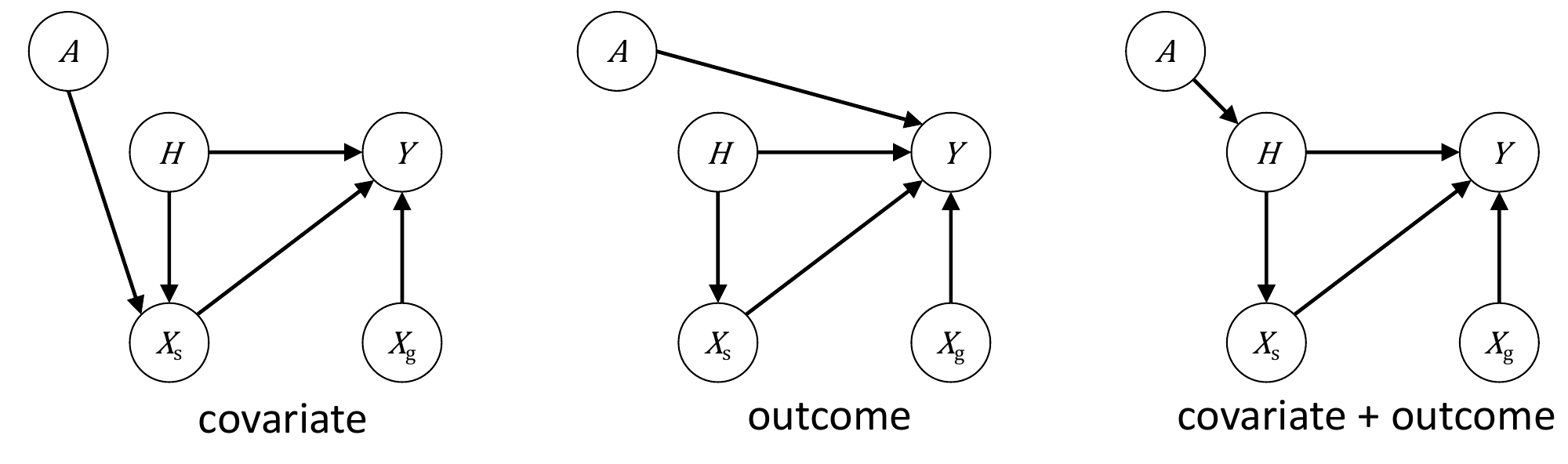}
\caption{Directed acyclic graphs (DAGs) illustrating the influence of a subgroup-specific shift variable $A$ on an outcome $Y$, covariates $X_s$, or both. Scenario \textit{covariate}: The shift $A$ affects only a subset of covariates $X_s$, with $H$ as a common confounder, maintaining a stable $\text{P}(Y|X)$ across subgroups. Scenario \textit{outcome}: $A$ directly affects $Y$, causing outcome shifts without influencing $X$. Scenario \textit{covariate + outcome}: A latent variable $H$, influenced by $A$, affects both $X_s$ and $Y$. $X_{\text{g}}$ are covariates independent of the subgroup.}
\label{scm_fig}
\end{center}
\end{figure}

When studying differences between data sets or subgroups, these can occur with respect to different parts of the underlying structure, which in turn can influence the performance of a subgroup-specific prediction model. Specifically, we introduce three scenarios (see Figure \ref{scm_fig}), where subgroup-specific differences can either occur on the covariates, the outcome, or both.
As a tool for assessing the potential benefit of leveraging data from external subgroups, we focus on the increase in the effective sample size (ESS) achieved through the proposed weighting approach. This is compared to using only the sample size of the target subgroup. An increase in ESS indicates that our method is successfully incorporating data from external subgroups, thereby enriching the prediction model and potentially enhancing its prediction performance.

In Section 2, we introduce the different scenarios and describe the proposed weighting approach. Section 3 introduces a simulation study, in which we assess performance in settings with various degrees of similarity between subgroups. Section 4 shows an exemplary clinical application. Finally, we discuss our findings and explore potential extensions in Section 5.

\section{Methods}

\subsection{Estimating similarity of individual observations through propensity scores}

In our analysis involving multiple external subgroups and a specific target subgroup, we estimate the similarity of individual observations to the target subgroup using propensity scores. These scores are estimated for each combination of an external subgroup and the target subgroup subsequently. As an example, we define the propensity score estimation for one combination of an external subgroup and the target subgroup, where $\mathbf{s}$ is a binary vector with sample size $n$. An observation from the target subgroup is denoted as $s_j = 1$ and $s_j = 0$ for an observation from an external subgroup with $j \in (1,\ldots,n)$. For each comparison, we define the data matrix $\mathbf{Z}$ consisting of covariate matrix $\mathbf{X}$ and the outcome vector $\mathbf{y}$ with data from both the target and the external subgroup. The logistic regression formula with the resulting propensity scores $\hat{p}_j$ are given by:

\begin{equation}
\hat{p}_{j} = \text{P}(s_j = 1 | \mathbf{Z}_j) = \frac{\exp(\boldsymbol{\hat{\beta}}^T z_{j})}{1 + \exp(\boldsymbol{\hat{\beta}}^T z_{j})}
\label{eq:logistic}
\end{equation}

 where $\boldsymbol{\hat{\beta}}$ is the coefficient vector. Given our focus on deriving ideal weights for each subgroup, a logistic regression is employed to estimate the probability of an observation belonging to the target subgroup within each comparison, favoring binary classifications over the multinomial outcomes typically used in models like the one discussed in \cite{madjarWeightedCoxRegression2021}.

A key aspect of our method is the inclusion of the outcome vector $\mathbf{y}$ in the propensity score model. This approach enables us to capture subgroup-specific shifts in the data that affect either the covariates, the outcome, or both illustrated by the three scenarios in Figure \ref{scm_fig} (\textit{covariate}, \textit{outcome}, and \textit{covariate + outcome}. The estimated propensity scores $\hat{p}_{j}$ represent the first component of our weighting approach, measuring the within-subgroup similarity and accounting for variance within each subgroup's data. In Figure \ref{scm_fig}, this is introduced by the random variable $H$ representing an unobserved confounder.

\subsection{Adjusting propensity scores by inverse probability of subgroup similarity}

In the second component of our weighting method, we focus on quantifying subgroup-level similarity. This aspect is represented by variable $A$ in Figure \ref{scm_fig}, which indicates specific distribution shifts for each subgroup which can influence different aspects of the data. To reflect this in our weighting approach, we adjust the individual propensity scores $\hat{p}_j$ based on the subgroup similarity, effectively employing a type of inverse probability weighting.
Specifically, for external subgroups, we multiply the propensity scores by the inverse of the probability that the logistic regression model (from Equation (\ref{eq:logistic})) assigns a higher probability of belonging to the target subgroup for a randomly chosen observation from the target subgroup ($\hat{p}^{\text{target}}$) compared to one from the external subgroup ($\hat{p}^{\text{external}}$). As a consequence, all observations from a similar external subgroup are increased while still allowing for differences within each subgroup. In dissimilar external subgroups, our weighting approach still allows single observations to be similar to the target subgroup and receive a substantial weight. This inverse probability weighting is operationalized using the area under the curve (AUC) from the model. 
For the target subgroup, we assume a limited sample size $n_\text{target}$ and assign a weight of one to each observation, $w_{\text{target}} = \left( 1, \ldots, 1 \right) \in R^{n_\text{target}}$, ensuring its full representation in subsequent analyses. The definition of our weights is as follows:

\begin{equation*}
w_j  =
\begin{cases}
1 & \text{if observation $j$ is from target subgroup} \\
\hat{p}_j \times \frac{1}{\text{P}(\hat{p}^{\text{target}} > \hat{p}^{\text{external}} | s_j = 1, s_j = 0)} & \text{if observation $j$ is from external subgroup}
\end{cases}
\end{equation*}

for all individuals $j \in \left( 1, \ldots, n \right)$ and $w \in R^{n}$, where $n$ is the total sample size of the target and external subgroup. Further, $s_j = 1$ and $s_j = 0$ denote an observation from the target subgroup and an external subgroup, respectively.

In our weighting approach, we effectively address the various ways in which subgroups can differ as illustrated in Figure \ref{scm_fig}. Our method incorporates two key components: individual propensity scores to account for differences within subgroups, and an inverse probability weighting to quantify subgroup-level similarities. 

In situations with markedly dissimilar external subgroups, we suggest truncating weights to stabilize prediction performance. We define a threshold within the overlap between external and target propensity scores and set $w_j$ below this threshold to zero. This effectively excludes external observations lacking correspondence in the target data set. The threshold can be operationalized using a percentile, for instance, only including weights above the 5th percentile of the target propensity score distribution.

\subsection{Simulate distribution shift scenarios for multiple subgroups}

In our subgroup-specific prediction task with multiple subgroups, we assume that the underlying data generation process stems from a structural equation model (SEM) with latent variables \citep{bollen1989structural}, sometimes called structural causal model (SCM) \citep{pearlCausalInferenceStatistics2009}. An SCM is particularly useful in understanding and modeling the different similarities that exist both within and between subgroups. Further, the differences between subgroups can affect either the covariates only, the outcome only, or both the covariates and the outcome. We present three different simulation scenarios, illustrated in Figure \ref{scm_fig}, which draw inspiration from \cite{rothenhauslerAnchorRegressionHeterogeneous2021}. Each scenario results in a different type of distribution shift within the SCM, offering insights into how such shifts can influence the performance of subgroup-specific prediction models.

Based on the scenarios illustrated in Figure \ref{scm_fig}, we designed a simulation study to evaluate our proposed method. The three scenarios are distinguished by the effect of a subgroup-specific distribution shift value in $\mathbf{v}$, where $\mathbf{v} = [v_1,\ldots,v_m]$, on the variables of the SCM. Given $m$ distinct subgroups, each $a_j$ takes a value from $\mathbf{v}$ depending on subgroup membership where $a_j \in (1,\ldots,n)$. We apply no distribution shift to the target subgroup while shifting the distribution of several external subgroups. The extent of the distribution shift varies depending on the number of external subgroups, which we set at 1, 3, 5, or 7. To categorize the subgroups based on their similarity to the target subgroup, we introduce three levels of similarity: \textit{dissimilar} subgroups experience the most significant shift, up to a maximum of $v_m = 3$; \textit{medium similarity} subgroups undergo a moderate shift of $v_m = 2$; and \textit{similar} subgroups have the least shift with a maximum of $v_m = 1$. This categorization refers to the sampled data overall and describes the range of distribution shifts found among the individual subgroups. The specific shift for each subgroup within these categories is determined by a uniform sequence that starts at 0 for the target subgroup and incrementally increases up to the defined maximum shift for that category. For instance, in a scenario with three external subgroups classified as \textit{similar}, the sequence of distribution shifts would be set as $\mathbf{v} = [0, 0.33, 0.67, 1]$. Here, the target subgroup receives no distribution shift, and the subsequent values represent gradual increases in the shift intensity for each of the external subgroups. This systematic approach allows us to simulate a range of subgroup similarities and evaluate our method to assign weights based on this range of similarities. Before describing each scenario in detail, we define the data all scenarios have in common:

\begin{equation*}
    \begin{aligned}
        \mathbf{a} & \sim \text{Categorical}\left(\frac{1}{m}, \ldots, \frac{1}{m}\right) \\
        \epsilon_{\mathbf{h}}, \epsilon_{\mathbf{x}_{\text{s},p}}, \epsilon_{\mathbf{x}_{\text{g},i}}, \epsilon_{\mathbf{y}} \text{ indep. } & \sim \mathcal{N}(0, 1) \\
        \mathbf{x}_{\text{g},i} & = \epsilon_{\mathbf{x}_{\text{g},i}}, \quad \text{for } i = 1, \ldots, c \\
    \end{aligned}
\end{equation*}

The \textit{covariate} scenario (see Figure \ref{scm_fig}) examines how a subgroup vector $\mathbf{a}$ influences covariates $\mathbf{x}_{\text{s},p}$. Here, $\mathbf{a}$ directly impacts a subset of covariates, denoted $\mathbf{x}_{\text{s},p}$, creating subgroup-specific shifts. In contrast, $\mathbf{x}_{\text{g},i}$ are covariates unaffected by $\mathbf{a}$. A vector $\mathbf{h}$ serves as a confounder for both covariates $\mathbf{x}_{\text{s},p}$ and the outcome $\mathbf{y}$. Since the influence of $\mathbf{a}$ is channeled only through $\mathbf{x}_{\text{s},p}$, the conditional distribution $\text{P}(Y|X_{s,k})$ remains stable both within and across subgroups. For the \textit{covariate} scenario, the remaining variables are defined as follows:

\begin{equation*}
    \begin{aligned}
        \mathbf{h} & = \epsilon_{\mathbf{h}} \\
        \mathbf{x}_{\text{s},p} & = \epsilon_{\mathbf{x}_{\text{s},p}} + \mathbf{h} + \mathbf{a}, \quad \text{for } p = 1, \ldots, k \\
        \mathbf{y} & = 1\sum_{i} \mathbf{x}_{\text{g},i} + 1\sum_{p} \mathbf{x}_{\text{s},p} + 2\mathbf{h} + \epsilon_{\mathbf{y}} 
    \end{aligned}
\end{equation*}

The \textit{outcome} scenario reveals that pooling data across different subgroups can introduce bias because the conditional relationship between covariates and outcomes varies per subgroup. In this scenario, a direct pathway exists from $\mathbf{a}$ to outcome vector $\mathbf{y}$, causing a distribution shift that directly affects the outcome. There is no distribution shift of any of the covariates $\mathbf{x}_{\text{s},p}$ and $\mathbf{x}_{\text{g},i}$. Hence, both types exert a consistent impact on $\mathbf{y}$. Contrary to the \textit{covariate} scenario, variations in data structure among subgroups affect the conditional distribution $\text{P}(Y|X_{\text{s},p})$. The scenario-specific variables of the \textit{outcome} scenario are defined as follows:

\begin{equation*}
    \begin{aligned}
        \mathbf{h} & = \epsilon_{\mathbf{h}} \\
        \mathbf{x}_{\text{s},p} & = \epsilon_{\mathbf{x}_s,p} + \mathbf{h}, \quad \text{for } p = 1, \ldots, k \\
        \mathbf{y} & = 1\sum_{i} \mathbf{x}_{\text{g},i} + 1\sum_{p} \mathbf{x}_{\text{s},p} + 2\mathbf{h} + \epsilon_{\mathbf{y}}  + \mathbf{a}
    \end{aligned}
\end{equation*}

The \textit{covariate + outcome} scenario demonstrates that pooling data across subgroups can lead to bias when distribution shifts are influenced by a vector $\mathbf{h}$ (representing an unobserved confounder). In this setup, changes in $\mathbf{a}$ impact $\mathbf{h}$, subsequently affecting both $\mathbf{x}_{\text{s},p}$ and $\mathbf{y}$. While $\mathbf{x}_{\text{s},p}$ displays a subgroup-specific effect on $\mathbf{y}$, the effects of $\mathbf{x}_{\text{g},i}$ remain consistent across all subgroups. As with the \textit{outcome} scenario, the differences among subgroups modify the conditional distribution $\text{P}(Y|X_{\text{s},p})$. The SCM of the \textit{covariate + outcome} scenario is defined as follows:

\begin{equation*}
    \begin{aligned}
        \mathbf{h} & = \epsilon_{\mathbf{h}} + \mathbf{a} \\
        \mathbf{x}_{\text{s},p} & = \epsilon_{\mathbf{x}_s,p} + \mathbf{h}, \quad \text{for } p = 1, \ldots, k \\
        \mathbf{y} & = 1\sum_{i} \mathbf{x}_{\text{g},i} + 1\sum_{p} \mathbf{x}_{\text{s},p} + 2\mathbf{h} + \epsilon_{\mathbf{y}}
    \end{aligned}
\end{equation*}

In our scenarios, the effect of the categorical subgroup variable $\mathbf{a}$ is added as a linear shift intervention on one of the vectors in the system ($\mathbf{x}_{\text{s},p}$, $\mathbf{y}$, or $\mathbf{h}$). In the literature, this type of intervention is referred to as "dependent" \citep{korbVarietiesCausalIntervention2004}, "parametric" \citep{eberhardtInterventionsCausalInference2007}, or "mechanism change" \citep{tianCausalDiscoveryChanges2013}. For each scenario, this distribution shift then propagates through the system on a different level resulting in distributional changes. Importantly, within each subgroup the conditional distributions $\text{P}(Y|X_{\text{s},p})$ remain constant whereas joint distributions $\text{P}(Y,X_{\text{s},p})$ and marginal distributions $\text{P}(Y)$ and $\text{P}(X_{\text{s},p})$ can change depending on the scenario. This differs from concepts like moderator variables \citep{sharmaIdentificationAnalysisModerator1981} or effect modification \citep{vanderweeleFourTypesEffect2007} where the conditional distribution or in other words the effect of $X$ on $Y$ can change. As our measure of similarity between subgroups is based on propensity scores, the proposed method is only able to detect distribution shifts as found in our scenarios and not changes in the conditional distributions of outcomes and predictors.

In our simulation study, we assess the performance of our \textit{weighted} sample (using the proposed weights $w_j$) against two conventional training samples for prediction modeling. First, a \textit{global} sample, consisting of pooled data from all subgroups (e.g., medical centers), is likely the most common approach in multi-center studies in practice. For the \textit{global} sample, the sample size is less of a concern but rather calibration towards the target subgroup. As a second comparator, we use a \textit{local} sample consisting only of the data from the target subgroup. In a benchmark comparing domain generalization methods, \cite{zhangEmpiricalFrameworkDomain2021} found that when sufficient data are available, a \textit{local} sample almost always outperformed the other methods in terms of prediction performance.
The key performance metric was the RMSE on an independently sampled test set of the target subgroup from the same distribution as the training set and a sample size of 100. Throughout our results, we use simple linear regression models for prediction as the method itself is not the focus of this work but rather the weighting of data. More complex prediction methods can be easily used when some kind of weighting is applicable.

The prediction performance in the three-way comparison is primarily influenced by three factors: the simulation scenario (see Figure \ref{scm_fig}), the similarity between the subgroups, and the effective sample size (ESS) ratio between the \textit{weighted} and \textit{local} samples. Subgroup similarity reflects the magnitude of the subgroup-specific shift in our SCM. The ESS ratio between the ESS of the \textit{weighted} sample ($ESS_{\text{weighted}}$) and the \textit{local} sample size ($ESS_{\text{local}}$) indicates the extent to which similar external data is incorporated into the \textit{weighted} sample. This ratio was varied by sampling all combinations of 1, 3, 5, or 7 external subgroups with each a sample size of either 10, 20, 30, 40, or 50. As the focus is on improving prediction for small data, we varied the sample size of the target subgroup to be between 10 and 20 observations and four covariates (3 subgroup-specific $x_{\text{s}}$; 1 global $x_{\text{g}}$) for training. For evaluating our computed weights as a measure of similarity (see Figure \ref{fig:weight_fig}), we also examined sample sizes up to 30 to show how the weights evolve beyond the small data scenarios. Varying all aforementioned factors adds up to 1440 different scenarios. Each scenario was sampled 100 times, leading to 144,000 simulation runs. The resulting ESS ratio is then rounded to the nearest integer.

The code to run all experiments, create the figures, and do the weighting is written in the R programming language \citep{rcoreteamLanguageEnvironmentStatistical2023} with version 4.3.1 and the following packages {\tt{dplyr}} (v1.1.3), {\tt{ggplot2}} (v3.4.3), {\tt{Hmisc}} (v5.1.1), and {\tt{Metrics}} (v0.1.4). For code access and to replicate our simulation findings, please visit our GitHub repository at https://github.com/max-jonasbehrens/simcow/.

\section{Simulation Results}

\subsection{Weights as a similarity measure}

\begin{figure}[htb]
\begin{center}
\includegraphics[width=\textwidth]{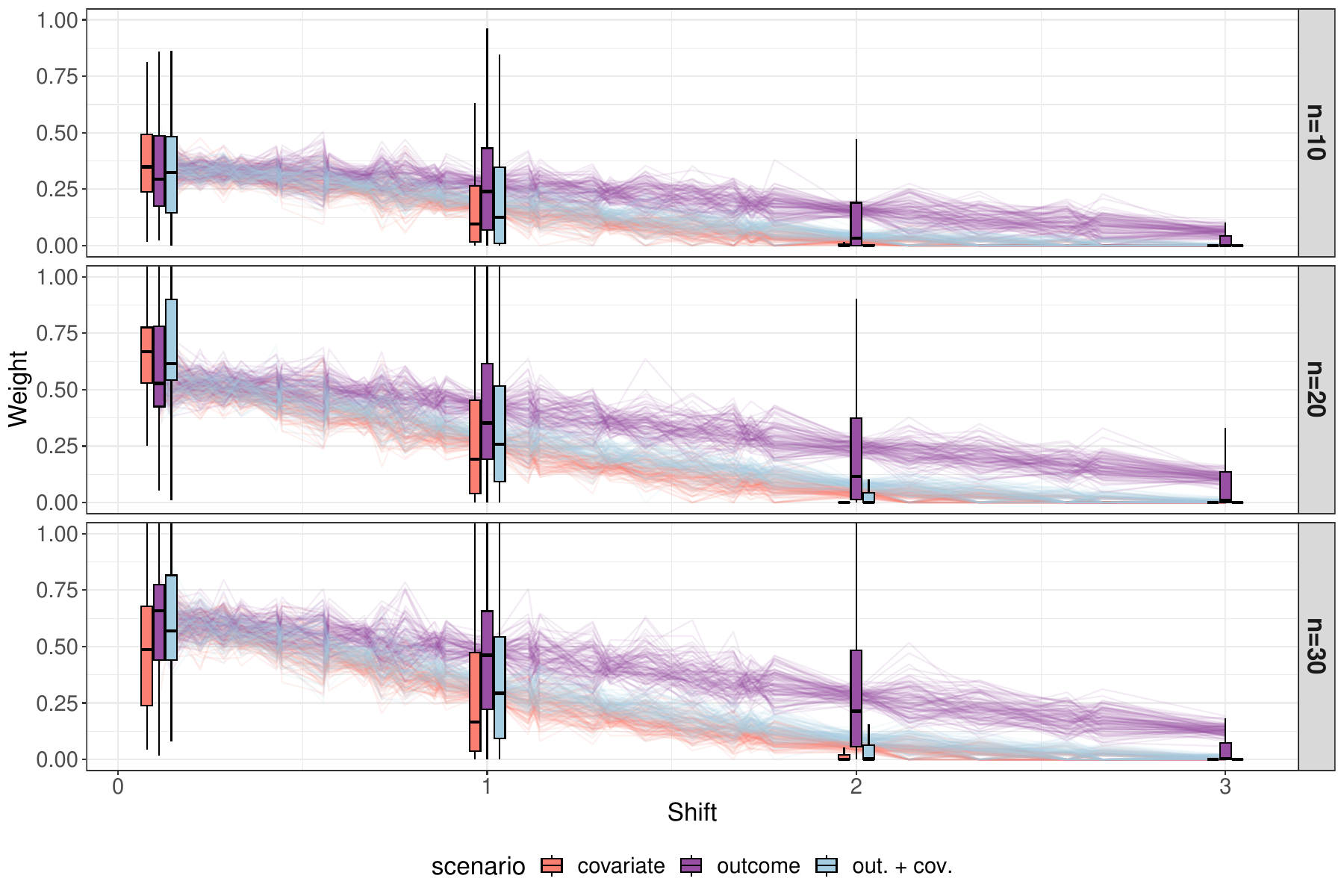}
\caption{Relationship between distribution shifts and computed weights in different simulation scenarios. Each row corresponds to the subgroup sample size (10, 20, 30). Within each panel, individual lines represent the averages for each subgroup subjected to distribution shifts (x-axis) across scenarios \textit{covariate} (orange), \textit{outcome} (purple), and \textit{covariate + outcome} (light blue). Notably, smaller shifts correlate with higher weights. The boxplots show the weight distribution within one example subgroup at specific distribution shifts (0.1, 1, 2, 3). This indicates the wide range of weights within each subgroup.}
\label{fig:weight_fig}
\end{center}
\end{figure}

Our approach relies on assigning weights to external data, which are determined based on their similarity to the target subgroup. There is a clear correlation between the calculated weights and the distribution shift applied to each external subgroup, as demonstrated in Figure \ref{fig:weight_fig}. This figure illustrates the relationship between the magnitude of the distribution shift (x-axis) and the corresponding weights (y-axis), across all shifts in our simulation. The lines represent the average weight calculated for each subgroup under these shifts. To illustrate the variation in weights within subgroups, boxplots are presented for an exemplary subgroup across distribution shifts of 0.1, 1, 2, and 3.

Figure \ref{fig:weight_fig} shows a negative correlation between the degree of distribution shift and the computed weights, consistent across all scenarios (\textit{covariate}, \textit{outcome}, \textit{covariate + outcome}) of our study. This trend is evident regardless of the sample size of the subgroups, which in turn influences the results in two ways. Larger sample sizes not only increase the average weights of a subgroup but also appear to reduce the variance of weights within a subgroup. Notably, in the \textit{outcome} scenario, where shifts impact only the outcome, even a distribution shift of 3 results in non-zero weights. Conversely, the \textit{covariate} and \textit{covariate + outcome} scenarios, which involve more complex shifts, yield lower weights, predominantly zero with a shift of 2 or higher. Across all scenarios and distribution shifts up to 2, a wide range of weights is observed within each subgroup which can be seen in the respective boxplots. The simulation study consists of the multivariate scenario with subgroup-specific ($x_{\text{s},p}$) and global covariates ($x_{\text{g},i}$). Larger samples enhance the accuracy of probability estimates, yet not all observations receive a weight close to 1. We tested our approach under a very simple scenario with one covariate and two large subgroups drawn from the same distribution. Here, weights of around 1 are achieved.

\subsection{Prediction performance in our simulation study}

\begin{figure}[htb]
\begin{center}
\includegraphics[width=\textwidth]{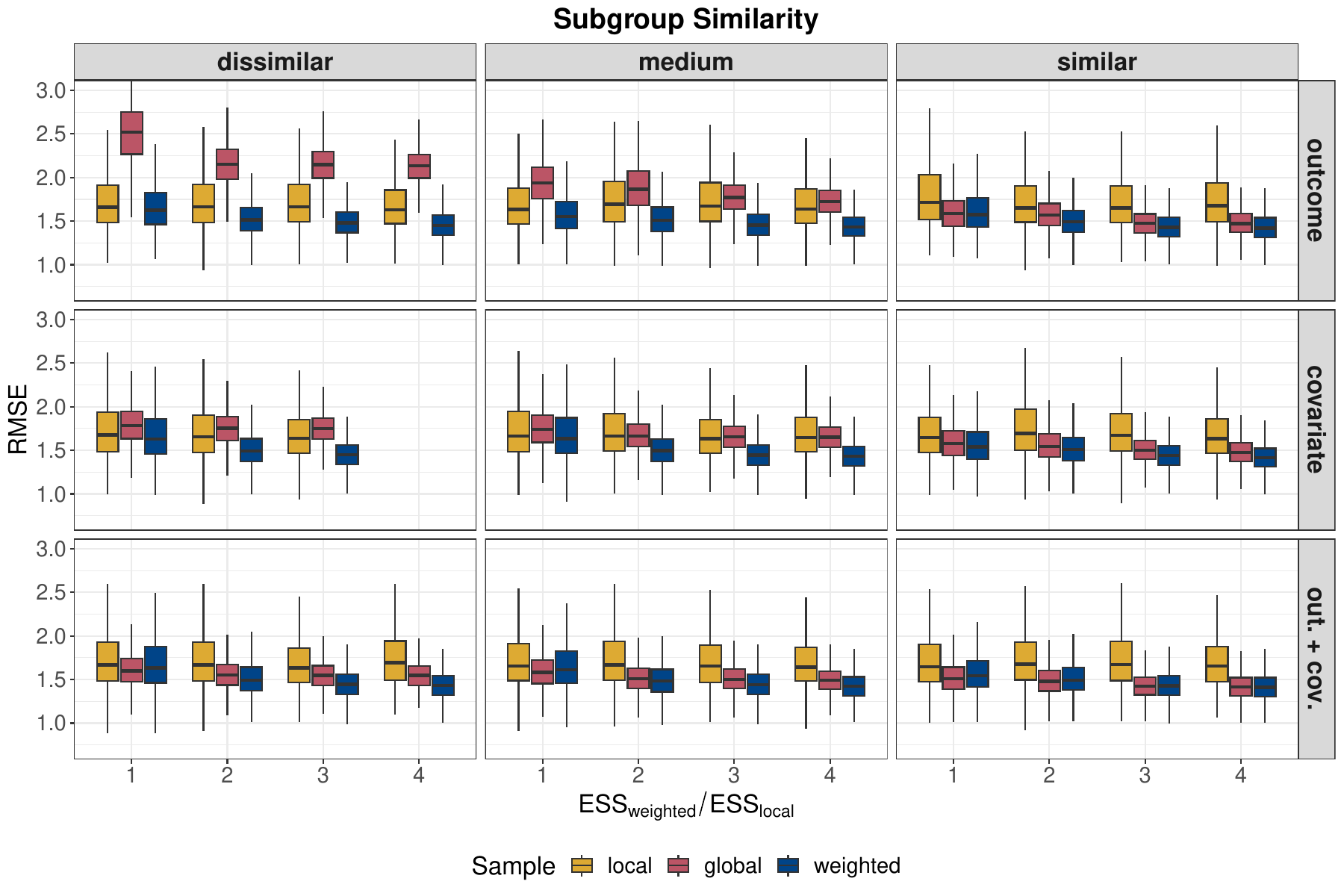}
\caption{Boxplot representation of the Root Mean Square Error (RMSE) across different subgroup similarity conditions (dissimilar, medium, similar) and sampling methods (global, local, weighted). Each panel corresponds to a distinct ESS ratio, ranging from 1 to 4. The RMSE values are stratified by the sampling strategy and are presented for three different cohorts, labeled \textit{outcome}, \textit{covariate}, and \textit{cov. + out} (\textit{covariate + outcome}), to illustrate the variation in prediction accuracy. The central line in each box represents the average RMSE, the box limits indicate the interquartile range (IQR), and the whiskers extend to the furthest points within 1.5 times the IQR from the hinge.}
\label{fig:sim_fig}
\end{center}
\end{figure}

Figure \ref{fig:sim_fig} illustrates the comparative prediction errors in terms of RMSE for the \textit{global}, \textit{local}, and \textit{weighted} samples, dissected by simulation scenario, subgroup similarity, and ESS ratio. The \textit{local} sample's performance remains relatively constant across scenarios, displaying high variance in prediction error due to limited observations (ranging from 10 to 20). Conversely, the \textit{global} sample's performance is dependent upon the scenario; it performs better in the \textit{covariate + outcome} scenario due to the larger sample size but shows a high bias in the \textit{outcome} and \textit{covariate} scenarios where the distribution shift is directly applied to the observed variables. This bias is mitigated when the similarity between subgroups increases. Within each grid of Figure \ref{fig:sim_fig}, the \textit{global} sample seems to benefit from an increasing ESS ratio, implying an influx of data more similar to the target subgroup.

Our \textit{weighted} sample generally surpasses both the \textit{global} and \textit{local} samples when similar external data is sufficient, with its superiority more pronounced in the \textit{outcome} and \textit{covariate} scenarios. The \textit{weighted} sample's RMSE and its variance tend to decrease as the ESS ratio climbs, suggesting that incorporating more similar data improves prediction accuracy. However, in highly dissimilar subgroups, the \textit{weighted} sample's ability to leverage external data is limited, which is evident in the lower observed ESS ratios. Overall, the \textit{weighted} sample shows superiority over the other samples, achieving the lowest prediction error when external data of some similarity is accessible.

Table \ref{tab_rank} compares the samples used for prediction across all simulation scenarios regarding the prediction error, and for the weighted samples the gain in ESS compared to the \textit{local} sample. We also run the simulations with the propensity score weights $\hat{p}_j$ only (see Equation (\ref{eq:logistic})) without the second component that adjusts for inter-subgroup similarities. Compared to our adjusted weights ($w_j$), the performance is somewhat decreased, indicating a potential benefit of a subgroup-specific weight component. Also, the difference in ESS ratio suggests that we gain on average around 15\% in ESS by adjusting the weights with the inverse probability weighting.

\begin{table}[htb]
\begin{center}
\caption{Comparing the samples throughout the simulation study.}
\label{tab_rank}
\begin{tabular}{lll}
\hline
Sample  & Average RMSE & Average ESS ratio \\
\hline
weighted & 1.527 & 2.852 \\
$\hat{p}_j$ only & 1.531 & 2.447 \\
local & 1.902 &  \\
global & 1.675 &  \\
\hline
\end{tabular}
\end{center}
\end{table}

\section{Clinical application results}

\subsection{A multi-center data set of patients with head and neck carcinoma}

Exemplarily, we will apply the proposed approach to predict the applied radiotherapy dose for elderly patients diagnosed with head and neck squamous cell carcinoma (HNSCC). This population has often been underrepresented in clinical trials and presents unique challenges due to increased comorbidities, frailty, and decreased organ function that can limit the usage of standard therapeutic options. The present patient population was part of an international cohort study known as SENIOR (NCT05337631) \citep{ruhleEvaluationConcomitantSystemic2023, ruhleMulticenterEvaluationDifferent2023}. The ethics committee of the University of Freiburg, Germany, approved this study in general, and the institutional review boards at each participating center approved data collection and data sharing with the responsible study center. This multi-center data presently consists of 1,100 older ($\geq$ 65 years) adults who are diagnosed with HNSCC and underwent definitive radiotherapy, either alone or with concomitant systemic treatment. Information about the patients and their treatments was gathered retrospectively from thirteen medical centers across the United States, Germany, Switzerland, and Cyprus. Our analysis primarily focuses on four covariates: the year of radiotherapy treatment, the Charlson comorbidity index (CCI), hemoglobin level, and leukocyte count, all measured before the commencement of radiotherapy treatment. These variables are utilized to predict the administered dose of radiotherapy.

The SENIOR study's heterogeneous data set, including varied patient profiles and treatment protocols from multiple medical centers, specifically focuses on those aged 75 years or older. Accurate prediction of the delivered dose of radiation is critical in this demographic due to their heightened sensitivity to treatment side effects and reduced tolerance levels. The challenge lies in the underrepresentation of this age group in existing data, combined with their distinct physiological responses and comorbidity profiles. Based on this demographic, five medical centers were identified with a suitable sample size (27, 24, 16, 26, and 64), showing center-specific differences to evaluate our weighting method. Our weighting method aims to enhance the accuracy of dose predictions for a target medical center by incorporating weighted data from other medical centers. The outcome of applied radiotherapy doses was not normally distributed. To address this non-normality, we apply a box-cox transformation with $\lambda = 2$, and square the radiotherapy dose values as a result. In Figure \ref{fig:senior_fig}, we revert these transformed values to their original scale. For the validation of our prediction model, we employ the .632+ Bootstrap method, as proposed by \cite{efronImprovementsCrossValidation6321997}. We run 1000 bootstrap samples to ensure the robustness and reliability of our model evaluation. 

\subsection{Prediction results of applied radiotherapy dose in the SENIOR data}

\begin{figure}[tbp]
\begin{center}
\includegraphics[width=0.85\textwidth]{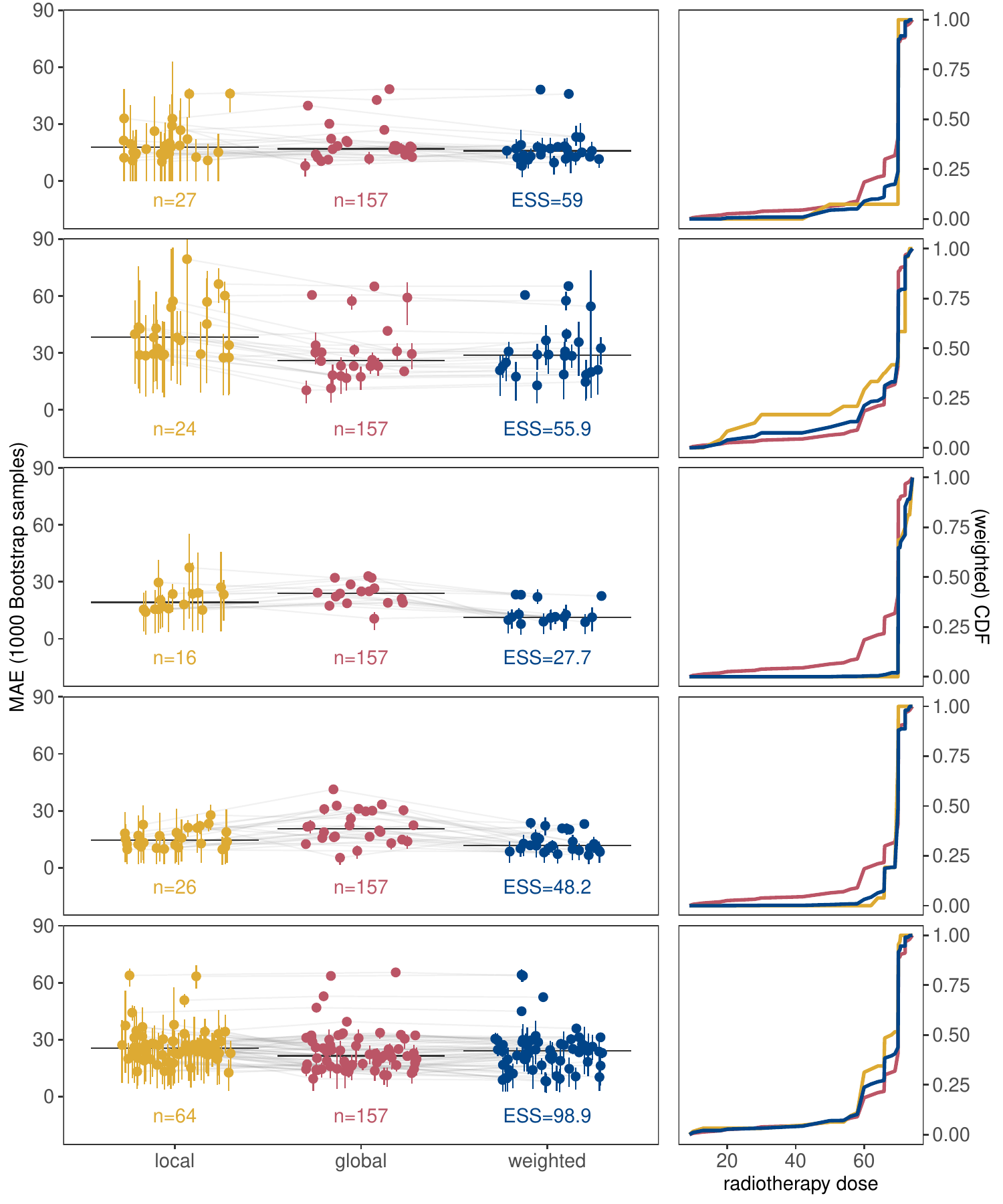}
\caption{Comparative prediction performance of \textit{weighted}, \textit{global}, and \textit{local} samples across multiple medical centers. Each row represents another target medical center, with the sample size (n or ESS) indicated for each model type. The left panels display scatterplots of the absolute prediction error obtained from 1000 bootstrap samples. Each dot within a column (the $x$-offset within each model column is uniform) corresponds to the mean absolute error (MAE) of 1000 bootstrap samples and the small bars indicate the 5th and 95th percentile of the absolute error. Identical observations are connected by grey lines and the overall median absolute error for each sample is indicated by horizontal bars. The right panels depict the corresponding (weighted) cumulative distribution functions (CDFs) for the radiotherapy dose.}
\label{fig:senior_fig}
\end{center}
\end{figure}

To put the method into practice, we applied it to the SENIOR data to predict the delivered radiation dose in this cohort. The results per medical center (in rows) are depicted in Figure \ref{fig:senior_fig}. The plot displays the distribution of the absolute prediction error per test set observation for the three different samples (\textit{weighted}, \textit{global}, and \textit{local}) and is based on 1000 Bootstrap samples. Aligned with the simulation results, the different samples are used to train a linear regression model as the underlying prediction model. Each plot indicates predictions for a different medical center with the sample size depicted in its respective color. On the right, we see the (weighted) CDF of the outcome variable to have an intuition about the similarity of the used samples.

The average prediction error of our \textit{weighted} sample consistently outperformed or matched the prediction error of the \textit{global} and \textit{local} samples across the depicted medical centers. Notably, this superior performance was achieved despite the \textit{weighted} sample's smaller ESS, underscoring our method's efficiency in selecting observations that enhance prediction accuracy. Generally, the variance of the prediction error distribution also seems to be lower for the \textit{weighted} sample. This might result from a combination of sufficient observations and a higher similarity toward the target medical center. On the contrary, the \textit{local} sample suffers from high variance of the prediction error and mostly inaccurate predictions compared to the other two samples which is most probably due to the low number of observations in the sample. This aligns with the results from the simulation study. The \textit{global} sample performs better than the \textit{local} sample in terms of prediction error and has less variance in its prediction error distribution. For the medical centers 2 and 5, the performance is on par with the one of our \textit{weighted} sample which is also reflected by the respective CDF plots on the right. Here, the \textit{weighted} CDF (in blue) is nearly identical to the \textit{global} CDF (in red). For medical center 2, the \textit{local} CDF (in yellow) seems to diverge from the other two CDFs which our proposed method is not able to reproduce or is not predictive for the given target medical center. The depicted CDFs for medical center 5 (last row) appear fairly identical, suggesting no apparent effect of this medical center on the received radiotherapy dose. This similarity in distribution explains the similar prediction error between the \textit{global} and \textit{weighted} samples. For the other medical centers, it seems that if there is a pronounced difference between the \textit{local} and the \textit{global} CDFs and this also reflects in the \textit{weighted} sample, then the \textit{weighted} sample outperforms the \textit{global} sample. The most pronounced example of this observation is medical center 3 where it seems that the delivered radiation dose is generally much more consistent than in the other medical centers.

Another observation from Figure \ref{fig:senior_fig} concerns the ESS ratio, the ESS of the \textit{weighted} sample by the ESS of the \textit{local} sample. The simulation results clearly show that the higher this ratio, the better the \textit{weighted} sample tends to perform. Based on the depicted sample sizes (n) or ESS in Figure \ref{fig:senior_fig}, the ESS ratio ranges from around 1.6 to 2.5 which is quite far from the upper end observed in the simulation results although the number of subgroups and sample size per subgroup were similar. This could indicate that it is more difficult for our proposed method to find similar data in the noisy application data and more data is needed to reach similar ESS ratios as seen in the simulation study. Thus, the prediction performance of the \textit{weighted} sample might further improve if more similar data were available.

\section{Discussion}

Clinical prediction models increasingly rely on observational data, which often comprises diverse subgroups, such as data from individual medical centers in multi-center studies. This diversity, influenced by factors like varying patient demographics and treatment protocols, poses significant challenges for building accurate subgroup-specific prediction models. Existing methods for subgroup-specific prediction modeling focus on one aspect of data similarity. Either they use subgroup-level weights to capture similarity between the subgroups or individual observation weights to quantify similarity to the target subgroup within an external subgroup. In three scenarios, we investigate differences that can occur between and within subgroups and can influence the covariates, the outcome, or both. Our proposed method combines both by providing individual weights and adjusting them according to the inter-subgroup similarity through inverse probability weighting. Further, we include covariates and outcome in the similarity quantification to capture all aspects of the data. To measure the amount of information our method can borrow from external subgroups, we look at the gain in ESS compared to the sample size of the target subgroup only.

Our method utilizes propensity scores to quantify similarities among individual observations. These scores are calculated using a logistic regression model that integrates both covariates and outcome data for each subgroup. The model estimates the probability that an observation in any given subgroup belongs to the target subgroup. The propensity scores, representing our first component, measure within-subgroup similarities. The second component of our approach quantifies subgroup-level similarity which involves adjusting individual propensity scores using inverse probability weighting. For external subgroups, we adjust these scores by the inverse probability that the logistic regression model assigns a higher probability to a randomly chosen observation from the target subgroup compared to a randomly chosen observation of an external subgroup. For the target subgroup with a limited sample size, we assign a weight of one to each observation, ensuring full representation in the analysis.

In the simulation study, we showed that our weighting method performs better compared to using only the target subgroup data or pooling the data across all subgroups. Throughout all scenarios, the weighting approach produces better prediction performance when sufficient similar external data is available (ESS ratio $\geq$ 2). By introducing more dissimilarity to the data, we saw that our weighting approach can struggle to find similar external data which in turn limits the capacity to improve predictions.

In a clinical application on the SENIOR data sets for predicting applied radiotherapy doses, our weighting approach outperformed or matched the model based on the pooled data as well as on the target data alone across the different medical centers. The effectiveness varied, with some centers showing only a slight improvement. Center-specific CDFs suggest that when the weighting approach captures the unique treatment pattern of a center, it predicts more accurately, particularly in cases where the global sample's bias is evident due to high dissimilarity. ESS ratios lower than those suggested by the simulation study, indicate a need for larger similar data sets in real-life applications to achieve improved predictions using our proposed weighting approach.

Since our weighting approach is based on a logistic regression model to calculate the weights, the method is limited in the type of differences between data sets or subgroups it can detect. As seen in our simulation study based on SCMs, the proposed weights can reflect linear distribution shifts that are directly or indirectly introduced to the observed variables. However, non-linear shifts or effect-modifying differences between subgroups will not be uncovered correctly by our method. As a result, some assumptions have to be made regarding the data at hand. Further, we show how to make good predictions in the context of small data. However, the small data problem extends to the underlying logistic regression model as well. If the target subgroup consists only of a few observations, our resulting weights might be unreliable in terms of the true similarity between subgroups.

Future research should tackle limitations related to non-linear distribution shifts and effect-modifying variations across different settings, enhancing our method's applicability. Key to this advancement is developing a more versatile approach for estimating similarities among subgroups, moving beyond logistic regression models. This endeavor is challenging, particularly with small datasets, which pose significant constraints. Our current method is designed for typical situations involving data stratified by categorical variables, such as medical centers. Future work could extend to scenarios involving continuous strata or a much larger range of categories than those considered in this study. A prime example is the analysis of time series data, where distribution shifts occur over time. In such cases, a new similarity measure must accurately capture temporal variations.

In conclusion, our method advances prediction modeling in small data scenarios by effectively utilizing similar external data, e.g., enhancing accuracy in diverse subgroups. By addressing the challenge of incorporating both differences between and within subgroups, our method offers a viable alternative for leveraging prediction modeling in, for instance, multi-center data. By looking at the gain in ESS, we show an effective way of understanding how our method finds similar external data. Lastly, our simulation introduces three scenarios of how we can systematically evaluate our method depending on the way subgroup-specific differences are affecting the data. Future developments should aim to extend its application to complex structures like time series, further broadening its impact in various prediction modeling contexts.
\vspace*{1pc}

\noindent {\bf{Acknowledgement}} This work was funded by the Deutsche Forschungsgemeinschaft (DFG, German Research Foundation) – Project-ID 499552394 – SFB 1597 (MF, AR, DZ, HB, and MB).
\vspace*{1pc}

\noindent {\bf{Conflict of Interest}}

\noindent {\it{The authors have declared no conflict of interest.}}

\bibliographystyle{elsarticle-harv} 
\bibliography{main}
\end{document}